\documentclass[sigconf]{acmart}
\AtBeginDocument{%
  }

\setcopyright{acmlicensed}
\copyrightyear{1234}
\acmYear{5678}
\acmDOI{}
\acmConference[]{CIDR}{Santa Cruz, California}{2026}
\acmISBN{xxxyyy}

\usepackage{xcolor}
\usepackage{balance}
\usepackage{colortbl}
\usepackage{adjustbox}
\usepackage{subfigure}
\usepackage{multirow}
\usepackage{pifont}
\usepackage{xspace}
\usepackage{enumitem}
\usepackage{bm}
\usepackage{tcolorbox}
\usepackage{subcaption}
\usepackage{makecell}

\newcommand{\stitle}[1]{\noindent\underline{\textbf{#1}}}

\newcommand{\ie}{\emph{i.e.}\xspace}
\newcommand{\eg}{\emph{e.g.}\xspace}
\newcommand{\wrt}{\emph{w.r.t.}\xspace}

\newcommand{\etc}{\emph{etc}}

\newcommand{\squishlist}{
	\begin{list}{$\bullet$}{
		\setlength{\itemsep}{0pt}
		\setlength{\parsep}{3pt}
		\setlength{\topsep}{3pt}
		\setlength{\partopsep}{0pt}
		\setlength{\leftmargin}{1.0em}
		\setlength{\labelwidth}{1em}
		\setlength{\labelsep}{0.5em}
   }
}

\newcommand{\squishenum}{
	
	\begin{list}{\usecounter{scount}}{
		\setlength{\itemsep}{0pt}
		\setlength{\parsep}{3pt}
		\setlength{\topsep}{3pt}
		\setlength{\partopsep}{0pt}
		\setlength{\leftmargin}{1.2em}
		\setlength{\labelwidth}{1em}
		\setlength{\labelsep}{0.5em}
	}
}

\newcommand{\squishend}{
	\end{list}
}

\newcommand{\eat}[1]{}

\newcommand{\sysname}{\textsf{BridgeScope}}
\newcommand{\sys}{\textsf{BridgeScope}\xspace}
\newcommand{\baselinename}{\textsf{PG-MCP}}
\newcommand{\baseline}{\textsf{PG-MCP}\xspace}
\newcommand{\birdext}{\textsf{BIRD-Ext}\xspace}
\newcommand{\nlml}{\textsf{NL2ML}\xspace}
\newcommand{\gpt}{GPT-4o\xspace}
\newcommand{\claude}{Claude-4\xspace}

\definecolor{mygrey}{RGB}{230,230,240}

\usepackage{xpatch}
\usepackage[frozencache,cachedir=minted-cache]{minted}
\xpatchcmd{\minted}{\VerbatimEnvironment}{\VerbatimEnvironment\let\itshape\relax}{}{}

\begin{document}

\title{\sysname: A Universal Toolkit for Bridging Large Language Models and Databases}

\author{Lianggui Weng$^{*}$, Dandan Liu$^{*}$, Rong Zhu$^{\#}$, Bolin Ding, Jingren Zhou \\
\textsf{\large Alibaba Group}
}
\email{{ lianggui.wlg,   ldd461759,   red.zr,   bolin.ding,   jingren.zhou }@alibaba-inc.com}

\renewcommand{\shortauthors}{Lianggui Weng, Dandan Liu, Rong Zhu, et al.}

\begin{abstract}
As large language models (LLMs) demonstrate increasingly powerful reasoning and orchestration capabilities, LLM-based agents are rapidly proliferating for complex data-related tasks. Despite this progress, the current design of how LLMs interact with databases exhibits critical limitations in usability, security, privilege management, and data transmission efficiency. 
To resolve these challenges, we introduce \sys, a universal toolkit bridging LLMs and databases through three key innovations. First, it modularizes SQL operations into fine-grained tools for context retrieval, CRUD execution, and ACID-compliant transaction management, enabling more precise and LLM-friendly functionality controls. Second, it aligns tool implementations with both database privileges and user security policies to steer LLMs away from unsafe or unauthorized operations, improving task execution efficiency while safeguarding database security. Third, it introduces a proxy mechanism for seamless inter-tool data transfer, bypassing LLM transmission bottlenecks. All of these designs are database-agnostic and can be transparently integrated with existing agent architectures. We also release an open-source implementation of \sys for PostgreSQL. Evaluations on two novel benchmarks demonstrate that \sys enables LLM agents to operate databases more effectively, reduces token usage by up to 80\% through improved security awareness, and uniquely supports data-intensive workflows beyond existing toolkits, establishing \sys as a robust foundation for next-generation intelligent data automation.




\end{abstract}
\maketitle

\footnotetext{*: Equal Contribution, $^{\#}$: Corresponding Author}

\vspace{-0.5em}
\section{Introduction}~\label{sec:introduction}
\vspace{-0.5em}


\stitle{Background.}
Large language models (LLMs) have been widely applied in data-related tasks, such as data manipulation, data cleaning, and exploratory data analysis~\cite{llm-data}, to enable more automated and intelligent solutions. Due to the intrinsic complexity of these tasks, they are typically decomposed into subtasks that are manageable for the LLM using expert-crafted frameworks~\cite{llm-data}. For instance, a typical LLM-based NL2SQL pipeline involves three subtasks, \ie, schema linking, SQL generation, and SQL revision~\cite{nl2sql-survey}. However, such decompositions are highly specific to task scenarios and lack generalizability to broader task classes. As the landscape of data-related tasks grows in both scale and complexity, relying solely on manually crafted frameworks becomes increasingly impractical.

\eat{Following the rapid progress of large language model (LLM)-enhanced natural language-to-SQL (NL2SQL) techniques, traditional code-heavy data-related tasks (\eg, data manipulation~\cite{}, data cleaning~\cite{}, data integration~\cite{}, and data analysis~\cite{}) are now increasingly approached with LLMs more intelligently. 

Current LLM-powered solutions for data-related tasks typically rely on fixed, expert-crafted~\cite{} or user-defined pipelines~\cite{}, where LLMs serve as individual components for semantic parsing, reasoning, or intent recognition. The long-standing research on classic tasks such as data visualization~\cite{} and insight generation~\cite{} has given rise to the development of well-established, standardized pipelines (\eg, first generating a visualization query with LLM, then rendering the plot) that achieve convincing performance. However, these pipelines are highly rigid and difficult to generalize to broader application scenarios. To handle data-related tasks more flexibly, systems such as \textsf{PALIMPZEST}~\cite{} allow users to declaratively specify their goals with LLM-powered operators. As the main drawback, this approach requires users to possess substantial domain expertise to construct and maintain effective workflows. 
}


Recently, advanced LLMs (\eg, GPT-4o, DeepSeek R1, and Claude-4) have demonstrated impressive capabilities in multi-step reasoning and task planning~\cite{llm-planning-survey}. In parallel, the introduction of the model context protocol (MCP)~\cite{mcp} has standardized LLM interactions with external systems, which greatly enhances LLMs’ abilities through access to broad domain-specific tools. These advances are together ushering in a new era of \emph{fully automated}, \emph{general-purpose} agents capable of flexibly orchestrating and executing diverse, particularly data-related, tasks with minimal human intervention.

\begin{figure}[t]
  \centering
  \includegraphics[width=0.9\linewidth]{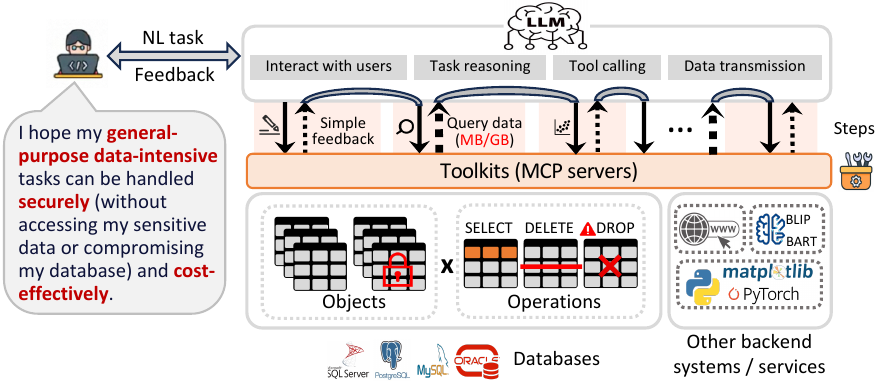}
  \vspace{-1em}
  \caption{Architecture of a general agent.}
  \label{fig:auto-reasoning}
  \vspace{-2em}
\end{figure}

As illustrated in Figure~\ref{fig:auto-reasoning}, such a \emph{general agent} employs an LLM as an intelligent manager, which: 
1) accepts and responds to users' natural language (NL) tasks; 
2) reasons over the task and orchestrates manageable steps to resolve the task (\eg, querying necessary data or performing data analysis), shown as pink-shaded blocks; 
3) invokes appropriate tools (black arrows) to handle each step; 
and 4) bridges and coordinates intermediate data flow from one tool to the next (possibly across backends). 
Towards this direction, both research~\cite{caesura, table-gpt} and production advances like Manus agent~\cite{manus} and ChatGPT agent~\cite{chatgptagent} have achieved state-of-the-art breakthroughs. 

\eat{
\stitle{General-purpose Frameworks for data-related Tasks.} 

In addition, the integration of tool-calling functionalities empowers LLMs to directly interact with databases (\eg, executing SQL queries) as well as with a wide range of external systems and services (\eg, retrieving data from the web or performing data analysis by writing Python scripts). 

Given a user's natural language request, the LLM can now act as an intelligent planner that automatically decomposes complex tasks into manageable sub-steps and invokes relevant tools to handle database operations (\eg, querying or manipulating data), as well as other processing or analytical tasks. Throughout the process, the LLM coordinates the data flow among various tools to ensure that the intermediate outputs are properly passed from one step to the next. Building on this paradigm, both research prototypes~\cite{caesura, table-gpt} and practical applications, exemplified by the general AI agent \emph{Manus}~\cite{}, have made large progress.

With the emergence of the Model Context Protocol (MCP)~\cite{mcp} that standardizes LLM's interaction with external systems, tool development is further decoupled from tool usage. This fosters a thriving open ecosystem where anyone can independently develop, share, and reuse tools. As a result, the range of tasks supported by the framework shown in Figure~\ref{fig:auto-reasoning} is greatly expanded, paving the way toward truly general-purpose systems.

\textsf{\emph{An Illustrative Use Case.}} Consider a common scenario in enterprise retail: a business analyst aims to revise product pricing strategies and develop a corresponding marketing plan based on historical sales data. As demonstrated in Figure~\ref{fig:auto-reasoning}, the analyst simply submits the overall goal in natural language. The planner LLM then decomposes the task into four steps: (S1) duplicating the current price strategy (to protect the original data); (S2) applying the new price strategy; (S3) collecting historical sales data relevant to price modifications; and (S4) generating a market plan and predicting potential revenue based on the curated sales data. In this pipeline, steps S1--S3 correspond respectively to database operations involving creating a duplicate price table, updating the price records, and selecting relevant sales data using join statements. The LLM handles these steps by invoking appropriate database tools. After that, the updated price tables from S2 and the sales data obtained in S3 are passed to advanced analytical tool-chains (\eg, training predictive models to estimate potential revenue under various marketing strategies) to accomplish S4, which ultimately delivers the final marketing recommendation and revenue projection. 

For example, a business analyst in a retail enterprise may wish to revise product pricing strategies and develop a new marketing plan. After the analyst submits this goal in natural language, the LLM might decompose the task into four subtasks: a) duplicating the current price table (to protect the original data); b) updating the new table to reflect the new price strategy; c) select historical sales data relevant to price modifications; and d) generating a market plan and predicting potential revenue based on the curated sales data. Steps S1--S3 correspond to database operations, and the LLM handles these steps by invoking database tools (\eg a \textsf{execute\_sql} tool that executes a given SQL statement). After that, the LLM passes the updated price tables from S2 and the sales data obtained in S3 to an advanced analytical tool (\eg, tasked with training predictive models to estimate potential revenue under various marketing strategies) to accomplish S4, which ultimately delivers the final marketing recommendation and revenue projection. 
}

\stitle{Challenges.} 
Current efforts in data-related scenarios focus on optimizing LLM-generated pipelines for complex tasks~\cite{caesura, table-gpt}. However, the orthogonal direction of designing database toolkits, which fundamentally shape the agents' capabilities in handling data, remains largely unexplored. Database toolkits within the MCP ecosystem~\cite{awesome-mcp-servers} are typically rudimentary, offering only a generic \textsf{execute\_sql} tool for executing SQL statements and a \textsf{get\_schema} tool for schema inspection. As analyzed below, such toolkits are inadequate for real-world data-related tasks, which in turn constrain the capabilities of agents and even expose the task-solving process to security risks, excessive resource consumption, and potential failures: 




$\bullet$ \textbf{C1. Coarse-Grained Tooling.} 
The universal \textsf{execute\_sql} tool is both insecure and cumbersome. On one hand, it grants agents unrestricted access to all user data and database operations, raising the risk of sensitive data exposure or unintentional destructive commands execution (\eg, \textsf{DROP DATABASE}) due to LLM hallucinations or prompt injection attacks~\cite{saastrai}. On the other hand, relying on a single tool for diverse operations, from querying data to managing transactions, greatly complicates LLMs' usage, increases their cognitive load, and heightens the risk of tool-selection errors. 





$\bullet$ \textbf{C2. Privilege-Unaware Planning.} 
Current agents lack intrinsic awareness of users' privileges in the database, and thereby may generate pipelines exceeding authorized operations (\eg, querying unauthorized tables). Although the database engine will eventually reject such operations, generating infeasible plans incurs non-negligible overhead for the LLM, especially when privilege violations occur late in the task execution workflow. 



$\bullet$ \textbf{C3. Direct Transmission of Voluminous Data.} 
The current agent architecture (Figure~\ref{fig:auto-reasoning}) that relies on LLMs to exchange data between tools is particularly problematic for data-intensive workflows. In this paradigm, the LLM's limited context window can be quickly exhausted and finally cause task failures. Furthermore, LLM hallucinations can introduce errors during transmission, further compromising the reliability.

\stitle{Our Contributions.} 
In response to these challenges (C1--C3), we present \sysname, a systematic database toolkit that improves both the effectiveness and efficiency of general-purpose LLM agents on data-related tasks. \textit{To the best of our knowledge, \sys is the first dedicated toolkit design in this domain}, establishing a foundation for future prototype and production-grade agent development.

At a higher level, \sys offers three core functionalities to support arbitrary data-related tasks: 
1) \emph{context retrieval}, which obtains task-related context (\eg, database schema and column exemplars) as essential background for LLM's interactions; 
2) \emph{SQL execution}, which enables general-purpose CRUD operations (\textsf{Create}, \textsf{Read}, \textsf{Update}, and \textsf{Delete});
and 3) \emph{transaction management} for end-to-end ACID-compliant task execution.
With each functionality implemented via a dedicated set of \emph{fine-grained} tools, \sys enables more precise and LLM-friendly control over database operations (as per Challenge C1). 
Instead of generic toolkits, \sys customizes tools aligning with both database-side user privileges and user-side security policies. Specifically, it includes only user-permitted objects with related privilege information in context retrieval results, and selectively exposes authorized, low-risk SQL execution tools (\eg, just the \textsf{select} tool for read-only users). This keeps LLMs aware of their operational boundaries, confines task planning to authorized scopes, and enables early identification and termination of infeasible tasks (addressing Challenge C2). Additionally, \sys employs rule-based verifications to strictly filter out unauthorized access and risky operations, providing an extra layer of protection. (see Sections~\ref{subsec:context-retrieval}--\ref{subsec:transaction} for details)


Beyond basic functionalities, \sys encapsulates a \emph{proxy} mechanism to support inter-tool data transfer (as per Challenge C3). With this approach, the LLM delegates data routing and execution management to a \textsf{proxy} tool, which retrieves necessary data and directly forwards it to downstream tools to trigger further operations, without any involvement from the LLM. This design not only mitigates the limitations of routing data through the LLM, but also allows seamless integration within the evolving MCP ecosystem, greatly expanding the agent's capabilities to a wider range of data-intensive tasks. (see Section~\ref{subsec:proxy} for details) 

\sys is database-agnostic and can be flexibly implemented for various database systems. We provide an open-source implementation of \sys for PostgreSQL~\cite{open-source-lib}. For rigorous evaluation, we introduce two novel benchmarks: one extends the NL2SQL benchmark BIRD~\cite{bird} to include complex database modifications, and the other involves training machine learning models using data extracted from the database. Experimental results show that \sys not only offers clear advantages in managing database operations such as transaction management, but also enhances LLM's security awareness, which contributes up to 80\% less token costs by intercepting infeasible tasks. Furthermore, \sys pioneers support for data-intensive workflows beyond the reach of existing toolkits. (see Section~\ref{sec:experiments} for details)




\vspace{-0.5em}
\section{\sysname: Design of the Toolkit}~\label{sec:sys-overview}
\vspace{-0.5em}

\sys abstracts essential functionalities from general data-related workflows and formulates core principles to guide tool design (Section~\ref{sec:func-abstraction}). For each functionality, a suite of fine-grained tools is then proposed (Sections~\ref{subsec:context-retrieval}--\ref{subsec:proxy}) to jointly address challenges C1–C3 (Section~\ref{sec:introduction}). Some remarks are provided in Section~\ref{subsec:implementations}. 



\vspace{-0.5em}
\subsection{Functionalities and Design Principles}~\label{sec:func-abstraction} 
\vspace{-0.5em}

\begin{figure}[t]
\centering
\hspace{1em}
\begin{minipage}[c]{0.33\linewidth}
    \centering
    \vspace{0.7em}
    \includegraphics[width=0.96\textwidth]{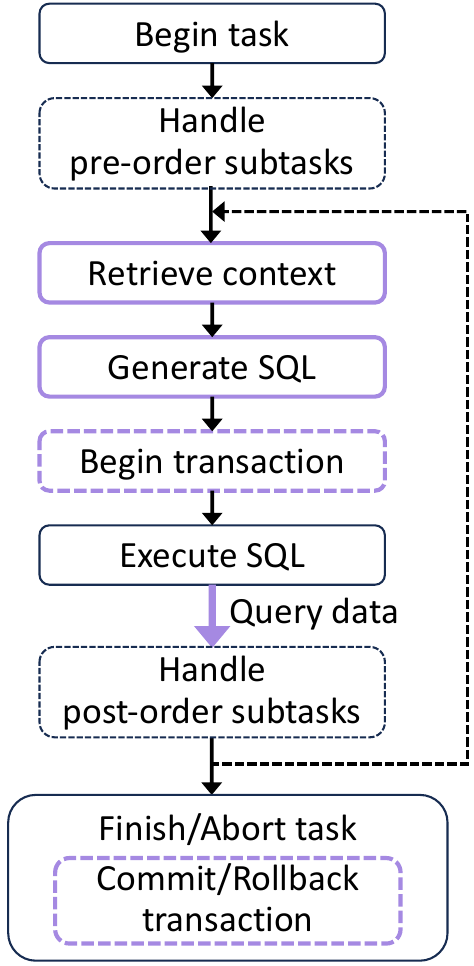}
\end{minipage}
\hspace{1em}
\begin{minipage}{0.57\linewidth}
\rule{\linewidth}{0.8pt}
\begin{minted}[tabsize=2,breaklines, fontsize=\scriptsize]{python}
# Tools for context retrieval:
def get_schema(): # return database schema
def get_object(o): # return object o's details
def get_value(col, key, k): # return top-k semantically relevant values w.r.t. task-specific key in column col's domain

# Tools for SQL execution:
def select(sql): # execute a SELECT SQL
def update(sql): # execute an UPDATE SQL
def insert(sql): # execute an INSERT SQL
... ...

# Tools for transaction management:
def begin(): # begin a new transaction
def commit(): # commit current transaction
def rollback(): # rollback current transaction

# Tool for data transmission:
def proxy(target_tool, tool_args): # execute target_tool with tool_args via proxy
\end{minted}
\vspace{-0.03in}
\noindent\rule{\linewidth}{0.8pt}
\vspace{-2em}
\end{minipage}
\caption{The abstracted workflow of data-related tasks (left) and fine-grained \sys toolkit (right).}
\label{fig:tool}
\vspace{-1.2em}
\end{figure}

\stitle{Functionalities Abstraction.}
We begin by examining key database operations in typical data-related workflows. As shown in Figure~\ref{fig:tool}(left), upon task submission, the LLM first handles all pre-order subtasks not requiring database access (\eg, user intent resolution). When data manipulation or querying is needed, it retrieves essential background information, such as the database schema, required to accurately formulate SQL statements (\textbf{F1. context retrieval}). It then generates and executes a valid, privilege-compliant SQL statement that fulfills the required goal (\textbf{F2. SQL execution}). For operations that modify the underlying database, the LLM must also initiate a transaction and execute these statements within the transaction context to ensure the ACID properties (\textbf{F3. transaction management}). After execution, feedback or resulting data are returned to the LLM and may be routed to follow-up tools (\textbf{F4. data transmission}) for subsequent processing or analysis~\cite{caesura}. If any step fails, all changes to the database are rolled back; otherwise, the workflow concludes with persisted database updates. Notably, one or more steps may iterate multiple times within a single task. 

For instance, in a chain store scenario where a Brand A manager daily updates sales/refunds records and analyzes recent sales trends, an LLM agent can facilitate this through a structured workflow: 1) retrieving relevant metadata such as table schemas and column names for sales and refunds records (F1); 2) atomically inserting daily records into the database (F2 and F3); 3) querying recent sales and refunds data from the database (F2); and 4) passing the results to an advanced machine learning tool (F4) for trend detection. 


\begin{figure}[t]
  \centering
  \includegraphics[width=0.95\linewidth]{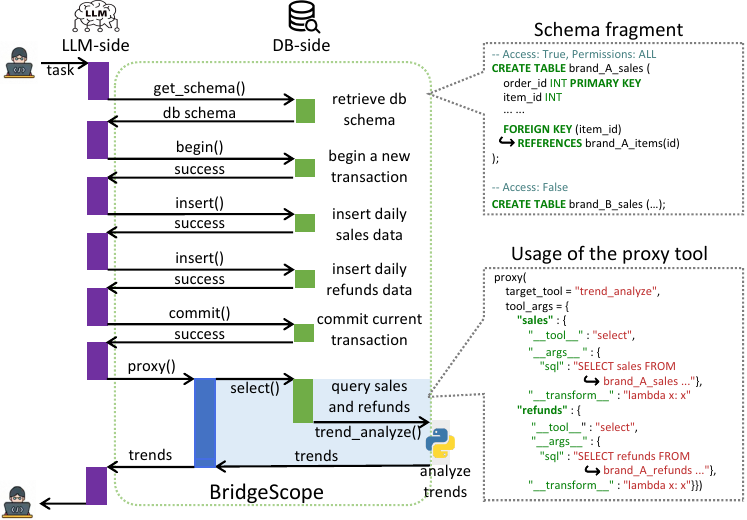}
  \vspace{-1em}
  \caption{An illustrative example of how \sys supports the chain store scenario.}
  \label{fig:example}
  \vspace{-1.5em}
\end{figure}

\stitle{Design Principles.}
From the above workflow and challenges C1--C3 stated in Section~\ref{sec:introduction}, we distill four key principles for tool design:

\textbf{1) Fine-Grained Tooling.} Rather than adopting a generic execution tool, tools should be fine-grained for specific LLM-database interactions to enable more flexible functionality controls and lower LLM's cognitive load for tool selection, as per Challenge C1. 


\textbf{2) Robust Security Guarantees.} 
Robust security must be enforced at dual levels: \emph{1) (\textbf{database-side})} Modern database systems like PostgreSQL and MySQL offer users granular privileges for actions (\eg, \textsf{SELECT}, \textsf{INSERT}, \textsf{UPDATE}, \textsf{DELETE}, and \etc) and objects (\eg, tables, views, columns, and \etc). Tools must ensure all operations adhere to these privileges; \emph{2) (\textbf{user-side})} Tools should support user-defined security policies to limit the LLM's privileges to a specific subset of the user's, thereby restricting its access to sensitive data and preventing destructive operations.


\textbf{3) High Efficiency.}
Toolkit design should prioritize efficiency in both time and computational resources, given the high cost of data-intensive tasks. It should empower the LLM to rapidly accomplish feasible tasks while promptly terminating infeasible ones. 

\textbf{4) Transparency.} Tools should operate transparently, neither interrupting nor requiring adaptations from other tools, thereby enabling seamless integration within the diverse and expanding MCP ecosystem to support a broad range of data-related tasks.

Guided by these principles, \sys implements each core functionality (F1--F4) using a set of dedicated tools, each corresponding to a specific database operation, as shown in Figure~\ref{fig:tool}(right). The rest of this section details the design of these tools and explains how they support the chain store task,  as demonstrated in Figure~\ref{fig:example}.


\vspace{-0.5em}
\subsection{Context Retrieval}~\label{subsec:context-retrieval}
\vspace{-0.5em}

\sys supports the retrieval of two key facets of task-related context: \emph{database schemas} and \emph{column exemplars}.


\stitle{Schema Retrieval.}
Schemas specify the structure and relationships among database objects (\eg, tables, columns, indexes, and foreign-key dependencies), which are crucial for accurate SQL generation~\cite{nl2sql-survey}. To support schema retrieval, \sys adopts an adaptive strategy to accommodate varying database scales: 


1) For databases with a manageable number of named objects (\ie, fewer than a user-specified threshold $n$), \sys provides a \textsf{get\_schema} tool that returns standardized, LLM-readable representations of all objects (see Figure~\ref{fig:example}). This enables the LLM to capture the complete database structure in a single tool call.

2) Otherwise, a hierarchical approach is adopted for more targeted and token-saving schema retrieval: \textsf{get\_schema()} returns only names of top-level objects (\eg, tables and views), while detailed information for a specific object $o$ (\eg, the columns, indexes, and constraints) are obtained by issuing \textsf{get\_object($o$)} as needed.



\sys employs two complementary mechanisms to steer LLM planning complying with the security requirements detailed in Section~\ref{sec:func-abstraction}. First, database-side privileges are made explicit to the LLM by augmenting schema outputs with privilege annotations for each object. As illustrated in Figure~\ref{fig:example}, the annotations above the table schemas indicate that the store manager has full access to the \texttt{brand\_A\_sales} table, but is restricted from \texttt{brand\_B\_sales}. More granular privileges (\eg, on specific columns) can be articulated similarly. 
Second, to limit LLM's access to sensitive data (\eg, the employee salary table) within the user’s privileges, \sys allows object-level restrictions from the user side via configurable white- and black-lists. The schema retrieval tools then expose only user-permitted objects to the LLM. Together, these mechanisms provide the LLM with clear awareness of access boundaries, guiding it toward compliant planning and less hallucinated access.

\stitle{Column Exemplars Retrieval.}
Due to synonyms, spelling variations, and domain-specific terms in database columns, LLMs often struggle to generate predicates aligned with actual data, leading to incomplete or inaccurate query results. By referencing column exemplars, LLMs can formulate semantically correct SQL predicates more accurately, \eg, generating predicate \textsf{category="women"} rather than \textsf{category="women's wear"} by examining the \textsf{category} column when analyzing sales trends for women's clothes. To support this, \sys provides a tool \textsf{get\_value(col, key, k)}, which retrieves the top-$k$ most semantically relevant values to a task-specific \textsf{key} (\eg, \textsf{"women"}) within column \textsf{col}'s domain. Notably, this targeted retrieval approach greatly reduces LLM context load compared to exhaustive enumeration of all column values.



\eat{
\begin{figure}[t]
\centering
\rule{\linewidth}{0.8pt}
\begin{minipage}{0.45\linewidth}
\begin{minted}[tabsize=2,breaklines, fontsize=\scriptsize]{sql}
-- Access:True, Permissions:ALL
CREATE TABLE brand_A_sales (
id TEXT PRIMARY KEY
date DATE
item_id INT
... ...
FOREIGN KEY (item_id) REFERENCES brand_A_items(id));
\end{minted}
\end{minipage}
\begin{minipage}{0.45\linewidth}
\begin{minted}[tabsize=2,breaklines, fontsize=\scriptsize]{sql}
-- Access:False
CREATE TABLE brand_B_sales (
id TEXT PRIMARY KEY
date DATE
item_id INT
... ...
FOREIGN KEY (item_id) REFERENCES brand_B_items(id));
\end{minted}
\end{minipage}
\rule{\linewidth}{0.8pt}
\vspace{-2em}
\caption{A schema fragment in \sys for the Brand A manager.}
\label{fig:get-schema-example}
\vspace{-2em}
\end{figure}
}

\vspace{-0.5em}
\subsection{Security-Gauranteed SQL Execution}~\label{subsec:sql-execution}
\vspace{-0.5em}




\sys adopts a two-level mechanism that enforces user-specific tool restrictions to help secure and efficient task execution.




\textbf{1) Action-Level Tool Modularization.} 
\sys modularizes SQL execution into a set of tools and selectively exposes them to the LLM based on both database-side privileges and user-side security configurations. Let the privilege set of user $u$ be $\mathcal{P}_u \subseteq \mathcal{A} \times \mathcal{O}$, where $\mathcal{A}$ denotes the set of database actions (\eg, \textsf{SELECT}, \textsf{INSERT}, \textsf{UPDATE}, \textsf{DELETE}, and \etc) and $\mathcal{O}$ denotes the set of database objects. For each possible action $a$, \sys instantiates a dedicated tool $T_a$ to exclusively handle SQL statements acting $a$ (\eg, the \textsf{select} tool for executing only \textsf{SELECT} statements). The agent for user $u$ is granted access to $T_a$ only if $(a, o) \in \mathcal{P}_u$ for some objects $o$. 
For instance, a chain store sales assistant with read-only access to the sales data receives only the \textsf{select} tool; while the manager is granted full CRUD tools, including the \textsf{insert} tool for daily sales records update. As with object-level restrictions (Section~\ref{subsec:context-retrieval}), users can further restrict LLM's tool access using white- or black-lists, \eg, by blocking the \textsf{drop} tool to prevent destructive actions like \textsf{DROP DATABASE}. These dual restrictions give the LLM an inherent understanding of its operational boundaries and effectively discourage its attempts to perform unauthorized or risky operations. 






\textbf{2) Object-Level Tool Verification.} 
Although \sys exposes permitted objects and privileges to the LLM via the \textsf{get\_schema} tool, access violations can still occur due to LLM hallucinations or prompt injections. To guarantee safe access, \sys enforces per-object verification during tool execution, ensuring that: 1) the user has privileges to operate the object; 2) the object complies with users' security policies. This tool-side check not only intercepts unauthorized operations, reducing the burden on the database, but also complements the database's native security mechanisms by additionally blocking dangerous actions specified by users. 

\vspace{-0.5em}
\subsection{Transaction Management}~\label{subsec:transaction}
\vspace{-0.5em}


\sys follows traditional database transaction controls and provides distinct \textsf{begin}, \textsf{commit}, and \textsf{rollback} tools to manage transactions. The ACID properties of each transaction (\ie, between \textsf{begin()} and \textsf{commit()}) are guaranteed inherently by the database engine. 
As Figure~\ref{fig:example} illustrates in the chain store scenario, the agent can invoke \textsf{begin()} to atomically insert both sales and refunds data, and call \textsf{commit()} to finalize the transaction upon success. Our experiments (Section~\ref{subsec:tooling-evaluation}) show that such explicit transaction tools substantially improve the LLM’s ability to manage transactions, resulting in more robust and reliable workflows.

\begin{figure}
    \centering
    \includegraphics[width=0.85\linewidth]{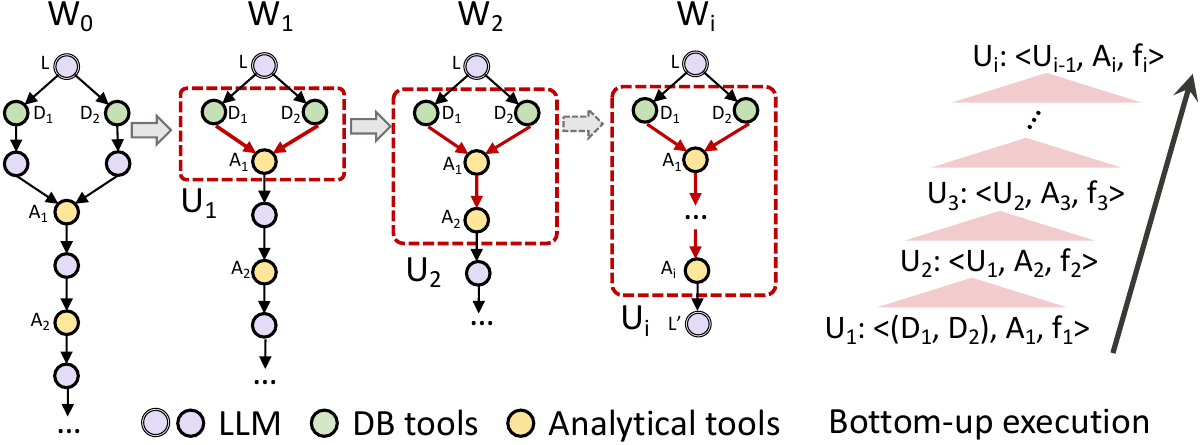}
    \vspace{-1em}
    \caption{Illustration of proxy units.}
    \label{fig:proxy-units}
    \vspace{-1.5em}
\end{figure}



\vspace{-0.5em}\subsection{Data Transmission with Proxy}~\label{subsec:proxy}
\vspace{-0.5em}



Data-related tasks usually involve a \emph{data producer} tool producing a large volume of data for use by a downstream \emph{data consumer} tool. To streamline data transfer, \sys introduces an advanced \emph{proxy} mechanism that automatically routes data between these tools, freeing the LLM from any involvement in transmission. 

\stitle{Foundations.} 
The proxy mechanism operates on each \emph{proxy unit} that specifies how data flows, formalized as a triple $\langle p, c, f \rangle$, where $p$ denotes a (or a list of) data producer(s) to collect necessary data, $c$ is the target data consumer, and $f$ is an adaptation function that transforms the outputs of $p$ to the input format expected by $c$. 

This mechanism supports arbitrarily complex data flow by allowing each proxy unit to act as a data producer for higher-level recursive proxy units. As Figure~\ref{fig:proxy-units} illustrates, in a typical data-related workflow without the proxy, as exemplified by $W_0$, where the LLM intermediates between each pair of tools to transfer data, the data flow from database tools $D_1$ and $D_2$ to analytical tool $A_1$ can be abstracted as a proxy unit $U_1 = \langle (D_1, D_2), A_1, f_1 \rangle$. Treating $U_1$ as a producer, its output to analytical tool $A_2$ further forms $U_2 = \langle U_1, A_2, f_2 \rangle$. This recursion continues until data transfer requires LLM-specific reasoning. In this example, the entire workflow between the initial and final LLM reasoning steps (\ie, $L$ and $L'$) can be abstracted as a hierarchical proxy unit $U_i$. 

During execution, the proxy units are processed in a bottom-up manner, with leaf units (\ie, $U_1$) handled first. Specifically, each data producer (\ie, $D_1$ and $D_2$) is invoked to collect necessary data, after which $f_1$ is applied and the transformed result is directly passed to the consumer $A_1$. The output of $A_1$ is then propagated upward in the proxy hierarchy to serve as input for the higher-level unit $U_2$. This process continues until the outermost proxy unit $U_i$ is processed and the result from $A_i$ is finally returned to the LLM ($L_i$).

\stitle{Benefits of Proxy.}
Data transfer in the execution of proxy units occurs directly between tools, bypassing LLM bottlenecks as detailed in Challenge 3 (Section~\ref{sec:introduction}). Beyond this core advantage, the proxy mechanism brings further practical benefits.


\textbf{1) Task Execution Efficiency.} 
By eliminating LLM-mediated data transfer steps, the proxy mechanism significantly streamlines task execution and shortens the critical path (\eg, condensing the workflow $W_0$ to the much simpler $W_i$ in Figure~\ref{fig:proxy-units}). Moreover, it enables parallel execution of multiple data producers (\eg, database tools $D_1$ and $D_2$), which further accelerates task completion.



\textbf{2) Adaptability.} 
By incorporating the function $f$ for automatic data format adaptation, the proxy mechanism operates transparently \wrt downstream tools. As highlighted in Section~\ref{sec:func-abstraction}, this is crucial for integration within the expansive MCP ecosystem and enables \sys to be seamlessly used with any domain-specific MCP servers to support diverse task scenarios. 




\stitle{Implementations.}
\sys implements the proxy mechanism via a \textsf{proxy} tool, complemented with a delicately crafted prompt that enables the LLM to recognize when to invoke it and autonomously generate proxy units as needed. 
The tool takes two arguments to define a proxy unit: \textsf{target\_tool}, which specifies the downstream tool $c$, and \textsf{tool\_args}, a dictionary that maps each input of \textsf{target\_tool} to its data producer(s) $p$ and a transformation function $f$ for format adaptation. When invoked, the proxy unit executes and the output of \textsf{target\_tool} is returned. This approach elevates the LLM from a passive data router to an active orchestrator of proxy units, while delegating data flow management entirely to the \textsf{proxy} tool.

Figure~\ref{fig:example} illustrates the use of the \textsf{proxy} tool for the chain store task. Here, \textsf{target\_tool} is set to a \textsf{trend\_analyze} tool, which analyzes sales trends via a machine learning model with lists of sales and refunds records (\textsf{sales} and \textsf{refunds}) as arguments. Each argument is produced by a \textsf{select} tool with an input \textsf{sql} querying recent sales or refunds data. As the results are already in the required format, an identity transformation is applied. Upon invocation, the \textsf{proxy} tool automates data selection, trend analysis, and inter-tool data transfer, and returns the sales trends produced by \textsf{trend\_analyze}.

\eat{
Here, each key in \textsf{tool\_args} corresponds to an input argument of the \textsf{target\_tool}, and its value specifies the database tool and its arguments for retrieving the input data. When invoked, the \textsf{proxy} tool prepares all inputs for \textsf{target\_tool} by calling corresponding database tools as defined in \textsf{tool\_args}, \eg, call "\textsf{select}(`SELECT x, y FROM table;')" to retrieve the input $P$ of \textsf{draw\_scatter}. After that, all resulting data is passed directly to \textsf{target\_tool}, triggering its execution. The details of this tool will be elaborated in Section~\ref{sec:proxy}. 

This tool takes two arguments: \textsf{target\_tool}, which specifies the downstream tool, and \textsf{tool\_args}, a JSON that describes how to obtain each input of \textsf{target\_tool} via database tools. 

For example, in a data visualization workflow, the target tool \textsf{draw\_scatter} takes a list $P$ of points as input and generates a corresponding scatter plot. \textsf{proxy} is used in the following way: 
\begin{minted}[tabsize=2,breaklines, fontsize=\small]{python}
    proxy(
        target_tool=draw_scatter, 
        tool_args = {
            "P": {
                "db_tool_name": "select",
                "db_tool_args": {
                    "sql": "SELECT x, y FROM table;"
                }
            }
        })
\end{minted}
Here, each key in \textsf{tool\_args} corresponds to an input argument of the \textsf{target\_tool}, and its value specifies the database tool and its arguments for retrieving the input data. When invoked, the \textsf{proxy} tool prepares all inputs for \textsf{target\_tool} by calling corresponding database tools as defined in \textsf{tool\_args}, \eg, call "\textsf{select}(`SELECT x, y FROM table;')" to retrieve the input $P$ of \textsf{draw\_scatter}. After that, all resulting data is passed directly to \textsf{target\_tool}, triggering its execution. The details of this tool will be elaborated in Section~\ref{sec:proxy}. 
}

\vspace{-0.7em}
\subsection{Remarks of \sys}~\label{subsec:implementations}
\vspace{-0.7em}

All tools in \sys are built on a unified set of database interfaces (\eg, executing a generic SQL) that can be implemented for any database system. This decouples tool logic from database specifics and makes migration to new databases straightforward. To showcase this, we release an open-source implementation for PostgreSQL~\cite{open-source-lib}. This database-agnostic design enables LLMs to interact with any data source using a consistent set of tools, without handling system-specific variations, greatly enhancing their capabilities in multi-datasource scenarios.



Our implementation also includes a carefully crafted prompt that enables more efficient and ACID-compliant LLM-database interactions. This prompt can be incorporated into any general-purpose agent to improve its handling of database-related steps. 
As shown in Section~\ref{sec:experiments}, the combination of \sys's toolkit and prompt achieves superior performance in supporting data-related tasks.






\vspace{-0.5em}
\section{Experimental Evaluation}~\label{sec:experiments}
\vspace{-0.5em}





This section comprehensively evaluates \sys to answer the following questions:
1) How does fine-grained tool modularization impact the handling of data-related tasks (Section~\ref{subsec:tooling-evaluation})?
2) How does privilege-aware tooling help intercept infeasible tasks and reduce unnecessary token costs (Section~\ref{subsec:privilege-evaluation})? 
and 3) What benefits can the proxy mechanism bring (Section~\ref{subsec:proxy-evaluation})? Notably, we tested scenarios involving privilege violations and operations exceeding users' security policies, all of which were successfully intercepted by \sysname’s rule-based security controls. Therefore, we omit further security evaluation in the following results.



\vspace{-0.5em}
\subsection{Experimental Setup}~\label{subsec:setup}
\vspace{-0.5em}

We implement two prototype general-purpose agents based on the ReAct~\cite{react} framework, with \gpt and \claude as the underlying LLMs, respectively. Throughout the evaluation, we refer to the two agents by their underlying LLMs. 

\stitle{Baseline.} We compare \sys with a baseline adapted from the official MCP server for PostgreSQL~\cite{awesome-mcp-servers}, referred to as \baseline. \baseline offers two tools: \textsf{get\_schema} for schema retrieving, and \textsf{execute\_sql} for SQL execution. This design is representative and widely adopted in current MCP servers for databases. 



\stitle{Benchmarks.} 
Existing benchmarks are too simple to reflect the full functional requirements of real-world data-related tasks. For rigorous evaluations of \sys, we synthesize the core requirements of these tasks and construct two novel benchmarks as below. Both benchmarks are made publicly available~\cite{open-source-lib}. 


\textbf{1) \birdext.} 
We extend the state-of-the-art NL2SQL benchmark \textsf{BIRD}~\cite{bird} originally focused on read-only \textsf{SELECT} SQL queries with complex data manipulation operations (\ie, \textsf{INSERT}, \textsf{UPDATE}, and \textsf{DELETE}). For each type of operation, we choose $50$ original \textsf{SELECT} tasks, modify the involved tables, and adapt the NL description to produce new tasks. The extended benchmark includes $150$ randomly sampled original \textsf{SELECT} tasks and $150$ newly synthesized tasks, which not only preserves the complexity and diversity of \textsf{BIRD}, but also adds stronger focuses on operation semantics, user privileges, and transaction management for data modifications.

\textbf{2) \nlml.} 
This benchmark simulates end-to-end model training on the California Housing dataset~\cite{house}, which has a single \texttt{house} table of 10 columns and 20,000 rows. It comprises 30 NL tasks at three complexity levels (10 tasks each): 1) basic data querying and model training, 2) additional data processing (\eg, normalization), and 3) further house price prediction, corresponding to one, two, and three layers of proxy unit abstraction, respectively. For example, the second level involves routing data through extraction, processing, and model training steps. Unlike prior benchmarks~\cite{nl2vis} focused on small-scale tasks, this benchmark reflects practical needs for large-scale data transfer and complex workflow orchestration. 









\vspace{-0.5em}
\subsection{Coarse-Grained vs. Fine-Grained Tooling}~\label{subsec:tooling-evaluation}
\vspace{-0.5em}




This experiment investigates how the granularity of tools for each core functionality of LLM-database interactions (as described in Section~\ref{sec:func-abstraction}) affects the handling of \birdext tasks. 

\begin{figure}[t]
\begin{minipage}{0.9\linewidth}
\centering
    \includegraphics[width=.8\textwidth]{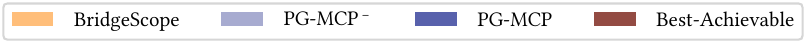} 
    \vspace{-0.5em}
\end{minipage}
\centering
\subfigure[Context retrieval]{
\begin{minipage}[c]{0.3\linewidth}
    \centering
    \vspace{-0.1em}
    \includegraphics[width=1\textwidth]{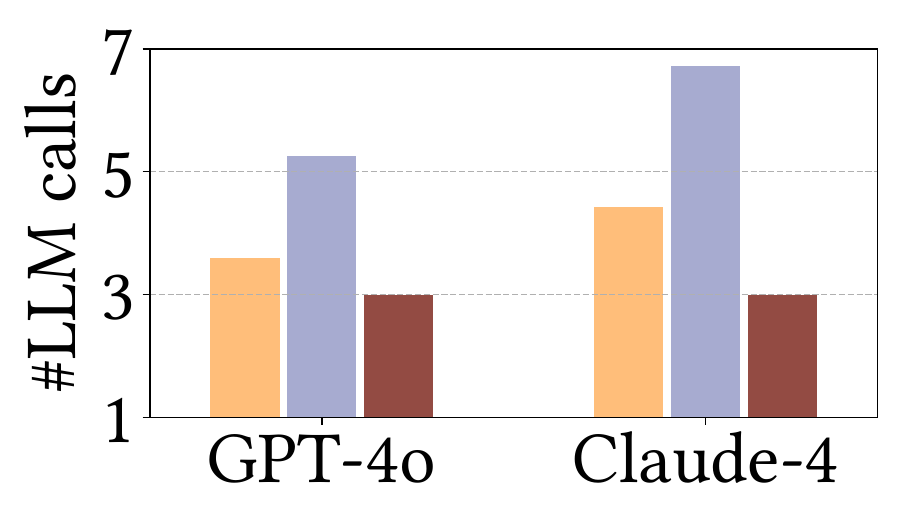}
\end{minipage}
}
\subfigure[SQL execution]{
\begin{minipage}[c]{0.3\linewidth}
    \centering
    \includegraphics[width=1\textwidth]{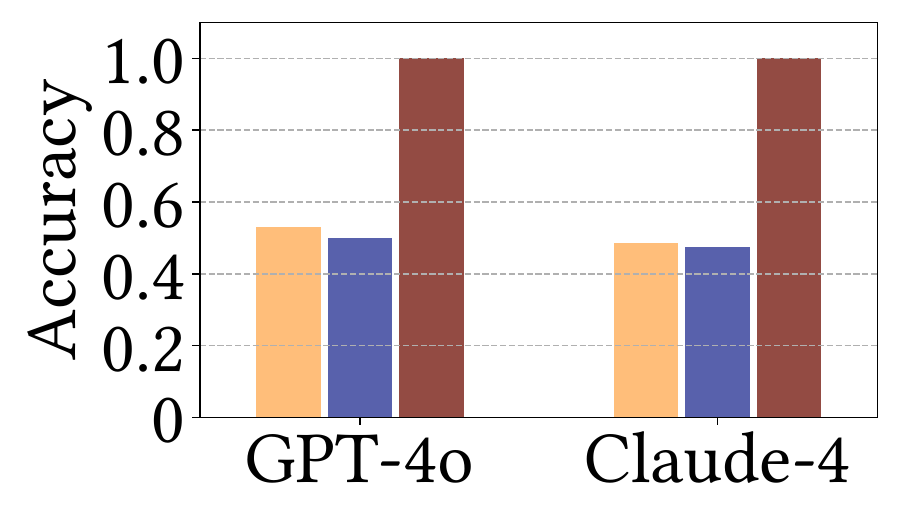}
\end{minipage}
}
\subfigure[Transaction management]{
\begin{minipage}[c]{0.325\linewidth}
    \centering
    \includegraphics[width=1\textwidth]{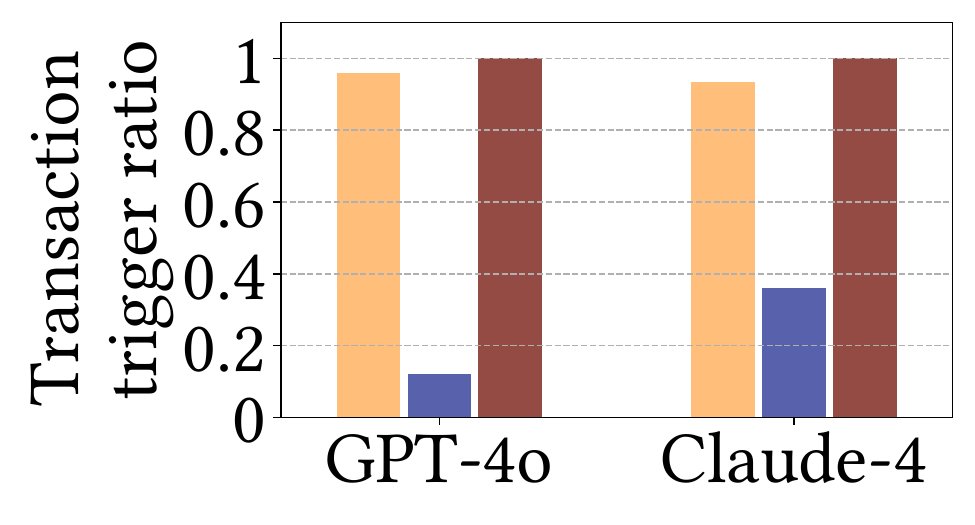}
\end{minipage}
}
\vspace{-1.5em}
\caption{Performance \wrt tooling granularity.}
\label{fig:fine-grained}
\vspace{-1.5em}
\end{figure}

\textbf{1) Context Retrieval.} 
To isolate the impact of explicit context retrieval tools, we compare \sys with a variant of \baseline that offers a single \textsf{execute\_sql} tool for both SQL execution and context retrieval, termed \baselinename$^-$. Figure~\ref{fig:fine-grained}(a) presents the average number of LLM calls they need to complete each task, which directly reflects the practical cost and efficiency. We observe that \sys reduces LLM calls by over 30\%, approaching the \emph{best-achievable} value of $3$ calls (one each for context retrieval, SQL execution, and result finalization). This gain arises because LLMs often hallucinate incorrect schema details and predicates with \baselinename-S, causing futile retries. In contrast, \sysname’s explicit context retrieval tools guide LLMs to gather necessary information beforehand, ensuring more reliable and efficient SQL generation.

 
\textbf{2) SQL Execution.} The main advantage of fine-grained SQL execution tools in \sys is their support for flexible security controls, with its benefits in execution efficiency later discussed in Section~\ref{subsec:privilege-evaluation}. Figure~\ref{fig:fine-grained}(b) compares the accuracy of agents handling \birdext tasks with \sysname’s fine-grained tools and \baselinename’s single \textsf{execute\_sql} tool. 
We observe that agents equipped with either \sys or \baseline achieve comparable accuracy on \birdext tasks. This indicates that the action-level modularization of SQL tools does not introduce unintended side effects on task completeness, thus confirming its practical viability. 



\textbf{3) Transaction Management.} 
The granularity of transaction management tools affects task control and ACID compliance, rather than efficiency or cost. Accordingly, we compare the ratio at which agents with \sys and \baseline correctly initiate transactions on \birdext tasks. Figure~\ref{fig:fine-grained}(c) demonstrates that agents with explicit transaction tools in \sys always initiate transactions correctly, with minor gaps to the theoretically best-achievable ratio of $1$ due to LLM's misjudgments to abort the tasks before SQL execution. In contrast, agents with \baseline rarely recognize the need for transaction management. This underscores the importance of surfacing key functionalities from generic execution tools to enhance the LLM's awareness of executing necessary operations. 




\vspace{-0.5em}
\subsection{Effectiveness of Privilege-Aware Tooling}~\label{subsec:privilege-evaluation}
\vspace{-0.5em}

To evaluate how \sys facilitates data-related tasks under different user privileges, we simulate three typical roles from production databases on the \birdext benchmark: \textbf{1) Administrator (A)}, with full data query and manipulation privileges; \textbf{2) Normal User (N)}, with read-only (\textsf{SELECT}) privileges; and \textbf{3) Irrelevant User (I)}, with privileges limited to task-unrelated tables. Tasks are categorized as either \textbf{read}, involving only data queries, or \textbf{write}, requiring data manipulation. 
Figure~\ref{fig:llm-calls-birdext} compares the average number of LLM calls required by \sys and \baseline to complete or, if infeasible for insufficient privileges, abort the tasks. Each bar corresponds to a specific combination of user and task types (\eg, \textsf{(A, read)} for an administrator performing a read task). Results for \textsf{(N, read)} are omitted as they closely resemble \textsf{(A, read)}. For reference, we also report the \emph{best-achievable} lower bound on LLM calls, as estimated in Section~\ref{subsec:tooling-evaluation} for feasible tasks, or as the minimum LLM steps needed to abort infeasible ones. Table~\ref{tbl:token-usage-bird-ext} summarizes the corresponding token costs. We have two main observations:

\begin{figure}[t]
\begin{minipage}{0.9\linewidth}
\centering
    \includegraphics[width=.6\textwidth]{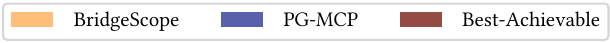} 
    \vspace{-0.8em}
\end{minipage}
\vspace{-1.2em}
\centering
\subfigure[Feasible tasks]{
\begin{minipage}[c]{0.035\linewidth}
    \centering
    \includegraphics[width=1\textwidth]{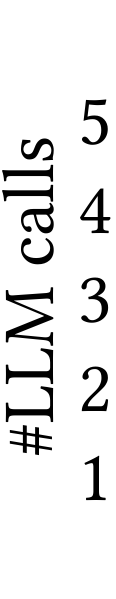}
\end{minipage}
\hspace{-0.5em}
\begin{minipage}[c]{0.18\linewidth}
    \centering
    \includegraphics[width=1\textwidth]{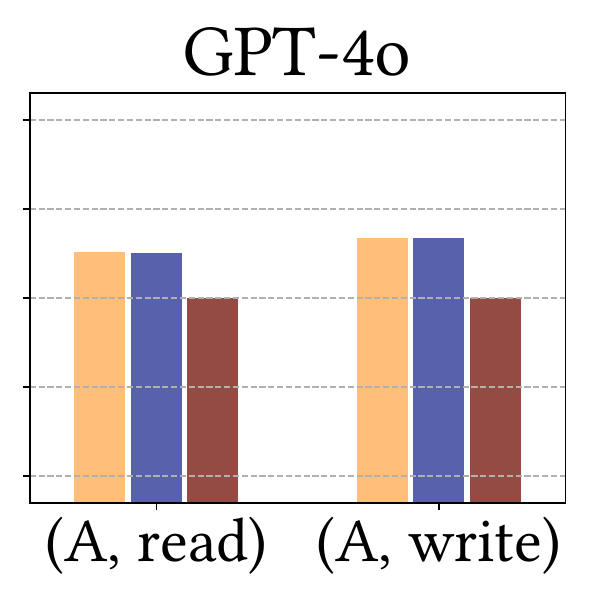}
\end{minipage}
\hspace{-0.7em}
\begin{minipage}[c]{0.18\linewidth}
    \centering
    \includegraphics[width=1\textwidth]{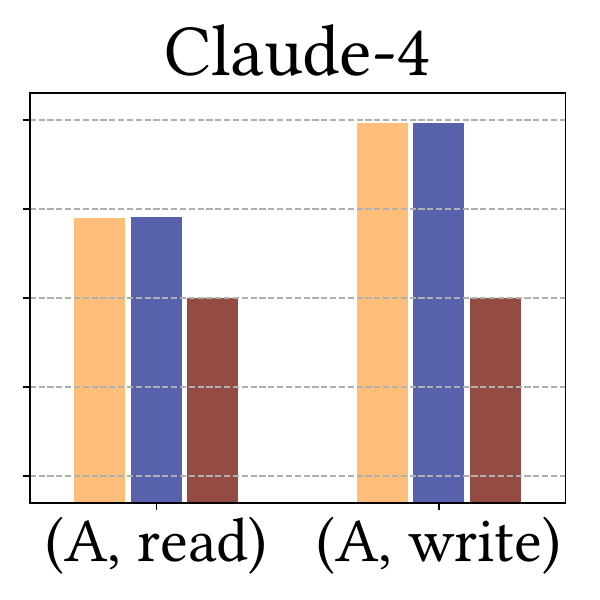}
\end{minipage}
}
\hspace{-0.5em}
\subfigure[Infeasible tasks]{
\begin{minipage}[c]{0.035\linewidth}
    \centering
    \includegraphics[width=1\textwidth]{figures/exp/y_label_permission.pdf}
\end{minipage}
\hspace{-0.5em}
\begin{minipage}[c]{0.27\linewidth}
    \centering
    \includegraphics[width=1\textwidth]{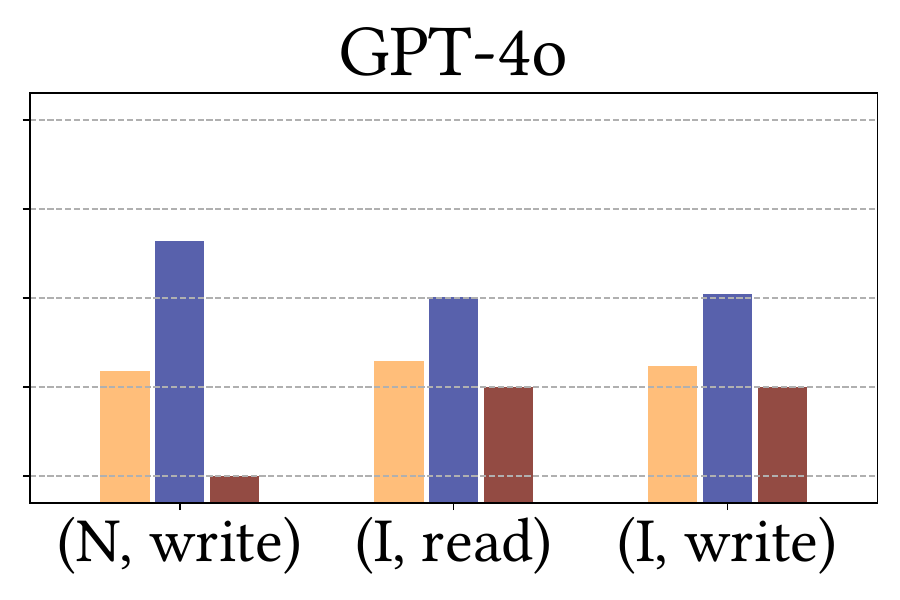}
\end{minipage}
\hspace{-0.6em}
\begin{minipage}[c]{0.27\linewidth}
    \centering
    \includegraphics[width=1\textwidth]{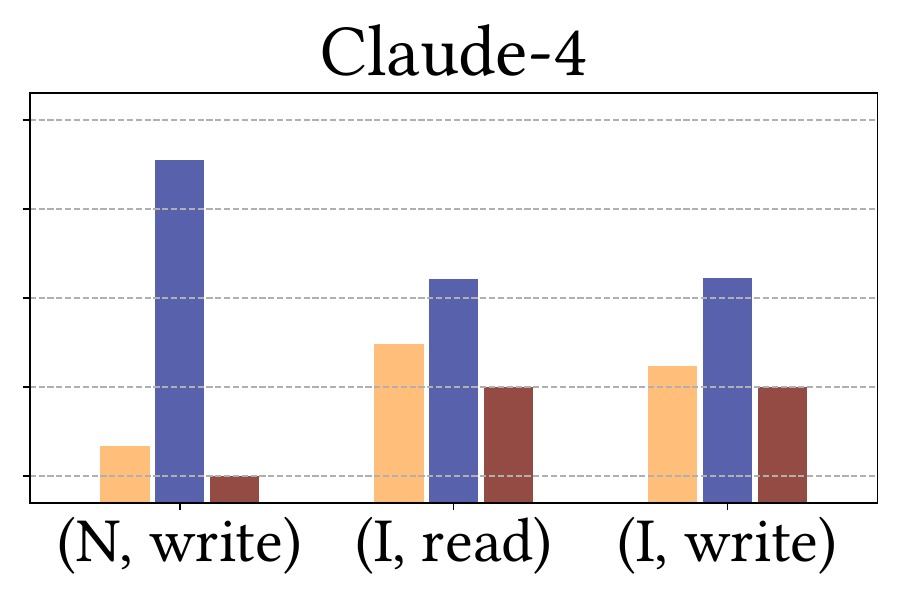}
\end{minipage}
}
\vspace{-0.5em}
\caption{Average number of LLM calls for \birdext.}
\label{fig:llm-calls-birdext}
\vspace{-1em}
\end{figure}

\renewcommand{\multirowsetup}{\centering}
\begin{table}[t]
    \centering
    \caption{Token usage for \birdext.}
     \vspace{-1em}
    \scalebox{0.7}{
    \begin{tabular}{|c|c|cc|ccc|} 
    
    \hline
    \rowcolor{mygrey}
    & & \multicolumn{2}{c|}{\textsf{Feasible tasks}} & \multicolumn{3}{c|}{\textsf{Infeasible tasks}} \\
    \cline{3-7}
    \rowcolor{mygrey} \multirow{-2}*{\textsf{Agent}} & \multirow{-2}*{\textsf{Toolkit}} & \textsf{(A, read)} & \textsf{(A, write)} & \textsf{(N, write)} & \textsf{(I, read)} & \textsf{(I, write)}   \\
    \hline
    \centering
    
    \multirow{2}*{\gpt} &
    \sys & 7,359 & 6,543 & 3,072 & 3,585 & 3,292 \\
    
    & \baseline & 7,282 & 6,746 & 6,391 & 5,124 & 5,091 \\
    \cline{1-7}
    \centering

    \multirow{2}*{\claude} &
    \sys & 10,102 & 14,652 & 2,194 & 4,515 & 4,632\\
    
    & \baseline & 9,803 & 15,111 & 12,109 & 6,496 & 7,093\\
    \hline
    \end{tabular}
    }
    \vspace{-0.5em}
    \label{tbl:token-usage-bird-ext}
\end{table}

1)  When users have sufficient privileges, \sys and \baseline incur similar numbers of LLM calls and token costs. This suggests that the privilege annotation and SQL tool modularization in \sys do not add cognitive burden for the LLM. 






2) For infeasible cases where users lack necessary privileges, \sys reduces the number of LLM reasoning steps by 23\%--71\%, with more pronounced improvements for \claude due to its stronger reasoning capabilities. These results approach the best-achievable bound, with the smallest gap of only 10\%. Correspondingly, token costs decrease by 30\%–82\%. These gains are attributed to \sysname's privilege annotation and SQL tool modularization, which help the LLM better understand user privileges and abort unattainable tasks before any SQL execution. The advantage is most evident when read-only users attempt write tasks (\ie, \textsf{(N, write)}). \sys exposes only the \textsf{select} tool to these users, allowing the LLM to promptly recognize its inability to manipulate data.

\begin{table}[t]
    \setlength{\tabcolsep}{1em}
    \centering
    \caption{Effectiveness of the proxy mechanism.
    }
     \vspace{-1em}
    \scalebox{0.7}{
    \begin{tabular}{|c|c|c|c|c|}
    
    \hline
    \rowcolor{mygrey}
    \textsf{Metric} & \textsf{Agent} & \sysname & \baselinename & \baselinename-S     \\
    \hline
    \centering
    \multirow{2}{*}{\makecell{Task Completion \\ Rate}} 
                  & \gpt & 1.0 & 0.0 & 1.0 \\
                  & \claude & 1.0 & 0.0 & 1.0 \\
    
    \hline
    \centering
    \multirow{2}{*}{\makecell{Token Usage \\ (On Average)}} 
                  & \gpt & 13,449.7 & - & 21,047.6 \\
                  & \claude & 15,622.3 & - & 22,353.1 \\
    
    \hline
    \centering
     \multirow{2}{*}{\makecell{\#LLM Calls \\ (On Average)}} 
                  & \gpt & 3.37 & - & 5.07 \\
                  & \claude & 3.40 & - & 5.07 \\
    
    \hline
    \end{tabular}
    }
     \vspace{-1.5em}
    \label{tbl:proxy}
\end{table} 

\vspace{-0.5em}
\subsection{Effectiveness of Proxy}~\label{subsec:proxy-evaluation} 
\vspace{-0.5em}

We evaluate the proxy mechanism on the \nlml benchmark. To handle these tasks, we equip agents with extra tools for data processing (\eg, Z-score normalization) and machine learning models (\eg, linear regression and random forest) training and inference. Table~\ref{tbl:proxy} compares \sys, \baseline, and its variant on three metrics: task completion rate (percentage of successfully completed tasks), token usage, and the number of LLM calls. We note that: 

1) \baseline struggles to complete any \nlml task because it relies on LLM’s limited context window to route data, which is quickly exhausted and causes task failures. In sharp contrast, \sysname’s proxy mechanism enables direct data routing between tools and allows \nlml tasks, even the most complex ones requiring three levels of proxy unit abstraction (see Section~\ref{subsec:setup}), to be completed with nearly the minimum possible number of three LLM calls (each for context retrieval, proxy execution, and result finalization). This demonstrates that modern LLMs can abstract complex proxy units and highlights the practical effectiveness of the proxy mechanism. 

2) Further examining \baselinename-S, a trivial version of \baseline that operates on a reduced \texttt{house} table with $20$ randomly sampled rows, we find that although it completes all \nlml tasks, 
its token usage and number of LLM calls still exceed those of \sys on the full table. This reveals that \baseline is extremely sensitive to data size and is only viable for very small data transfers, whereas \sys enables reliable task execution regardless of data size. 

3) For reference, we also consider \baseline incorporated in an idealized agent with unlimited reasoning capabilities and context window, which can, in principle, handle all \nlml tasks. However, the agent still incurs at least two transfers of the \texttt{house} table to handle each task, totaling no less than 1,500,000 tokens (750,000 per transfer). In contrast, \sys reduces token usage by more than two orders of magnitude (13,449.7 vs. $\ge$ 1,500,000), which could translate to significant cost savings. This result indicates that \sys can robustly handle large-scale data transfers with stable computational costs, highlighting its advantages for practical data-related tasks.

\vspace{-0.5em}
\section{Conclusions}~\label{sec:conclusion}
\vspace{-0.5em}

This paper presents \sys, a universal toolkit that bridges LLMs and diverse databases towards greater usability, efficiency, and security guarantees. Through fine-grained tool modularization, multi-level security controls, and a novel proxy-based data routing paradigm, \sys overcomes longstanding barriers in agentic database interactions. Evaluations on two novel benchmarks also confirm that, with \sys, LLM agents can handle data-related tasks with notably higher task completion rate and execution efficiency. All components in \sys are database-agnostic and transparent to both LLM agents and the MCP ecosystem, positioning \sys as a cornerstone for the next generation of trustworthy, scalable, and intelligent data-related automation.

\bibliographystyle{ACM-Reference-Format}
\bibliography{BridgeScope}

\end{document}